\documentclass[10pt]{article}
\usepackage{amsmath}
\usepackage{latexsym}
\usepackage{amsfonts}
\usepackage[normalem]{ulem}
\usepackage{soul}
\usepackage{array}
\usepackage{amssymb}
\usepackage{extarrows}
\usepackage{graphicx}
\usepackage{secdot}

\usepackage{subfig}
\usepackage{wrapfig}
\usepackage{wasysym}
\usepackage{enumitem}
\usepackage{adjustbox}
\usepackage{ragged2e}
\usepackage[svgnames,table]{xcolor}
\usepackage{tikz}
\usepackage{longtable}
\usepackage{changepage}
\usepackage{setspace}
\usepackage{hhline}
\usepackage{multicol}
\usepackage{tabto}
\usepackage{float}
\usepackage{multirow}
\usepackage{makecell}
\usepackage{fancyhdr}
\usepackage[toc,page]{appendix}
\usepackage[hidelinks]{hyperref}
\usetikzlibrary{shapes.symbols,shapes.geometric,shadows,arrows.meta}
\tikzset{>={Latex[width=1.5mm,length=2mm]}}
\usepackage{flowchart}\usepackage[paperheight=11.69in,paperwidth=8.27in,left=0.98in,right=0.79in,top=0.79in,bottom=0.79in,headheight=1in]{geometry}
\usepackage[utf8]{inputenc}
\usepackage[T2A]{fontenc}
\TabPositions{0.49in,0.98in,1.47in,1.96in,2.45in,2.94in,3.43in,3.92in,4.41in,4.9in,5.39in,5.88in,6.37in,}

\renewcommand{\_}{\kern-1.5pt\textunderscore\kern-1.5pt}

\usepackage{sectsty}

\sectionfont{\fontsize{11}{15}\selectfont}

\usepackage{indentfirst}

\setlistdepth{9}
\renewlist{enumerate}{enumerate}{9}
		\setlist[enumerate,1]{label=\arabic*)}
		\setlist[enumerate,2]{label=\alph*)}
		\setlist[enumerate,3]{label=(\roman*)}
		\setlist[enumerate,4]{label=(\arabic*)}
		\setlist[enumerate,5]{label=(\Alph*)}
		\setlist[enumerate,6]{label=(\Roman*)}
		\setlist[enumerate,7]{label=\arabic*}
		\setlist[enumerate,8]{label=\alph*}
		\setlist[enumerate,9]{label=\roman*}

\renewlist{itemize}{itemize}{9}
		\setlist[itemize]{label=$\cdot$}
		\setlist[itemize,1]{label=\textbullet}
		\setlist[itemize,2]{label=$\circ$}
		\setlist[itemize,3]{label=$\ast$}
		\setlist[itemize,4]{label=$\dagger$}
		\setlist[itemize,5]{label=$\triangleright$}
		\setlist[itemize,6]{label=$\bigstar$}
		\setlist[itemize,7]{label=$\blacklozenge$}
		\setlist[itemize,8]{label=$\prime$}

\usepackage{indentfirst}
\usepackage{lipsum}
\begin{document}
\setlength{\parskip}{0.0pt}
\begin{Center}
{\fontsize{14pt}{16.8pt}\selectfont \textbf{On coherent and incoherent scattering of fast charged particles in ultrathin crystals}\par}
\end{Center}\par

\vspace{\baselineskip}
\begin{Center}
\textbf{\textit{N.F. Shul'ga\textsuperscript{1,2}, V.D. Koriukina\textsuperscript{1}}}
\end{Center}\par

\begin{Center}
\textsuperscript{1}{\fontsize{10pt}{12.0pt}\selectfont  National Science Center "Kharkov Institute of Physics and Technology", 61108 Kharkov, Ukraine\par}
\end{Center}\par

\begin{Center}
\textsuperscript{2}{\fontsize{10pt}{12.0pt}\selectfont  Karazin Kharkov National University, 61000 Kharkov, Ukraine\par}
\end{Center}\par
\vspace{\baselineskip}

{\fontsize{10pt}{12.0pt}\selectfont We consider the fast charged particles scattering in ultrathin crystals on the base of the Born approximation of quantum electrodynamics. The main attention is paid to the question of the scattering cross section splitting into coherent and incoherent components when one of the crystallographic axes and planes is oriented along the direction of particle motion. It is shown that both the coherent and the incoherent components of the scattering cross section considerably depend on the orientation of the crystallographic axes relatively to the incident beam. In particular, it was shown that when particles are scattered by the crystal planes of atoms, the incoherent scattering cross section does not contain the Debye-Waller factor.\par}

{\fontsize{10pt}{12.0pt}\selectfont PACS: 03.65.Nk, 11.80.Fv\par}

\begin{multicols}{2}

\section {INTRODUCTION}

{\fontsize{10pt}{12.0pt}\selectfont When fast charged particles pass through matter, various coherent and interference effects are possible in their interaction with the atoms of the medium. The existence of such high-energy effects in ultrarelativistic electrons radiation in oriented crystals was noted in the work of Ter-Mikaelyan \cite{MLT53} and in an amorphous medium in the work of Landau and Pomeranchuk \cite{Lan}. Similar effects appear in other electromagnetic processes at high energies, such as the electron-positron pairs formation, ionization energy losses of particles in substances, etc. (see the monographs \cite{MLT72}\cite{Ahi96} and references in this them). Such effects are caused by interaction of particles with atoms of the medium within coherence lengths (formation lengths) of these processes, which at high energies can be macroscopic size.\par}

{\fontsize{10pt}{12.0pt}\selectfont Of particular interest is the process of fast charged particles scattering in thin layers of a substance, since the coherent and interference scattering effects in this case are the most considerable. Moreover, in some cases it is possible to develop methods and approximations to describe the scattering process in thin layers of matter, which significantly simplify the analysis of the scattering process. One of them is based on the consideration of the scattering process in the Born approximation of the quantum scattering theory. The present paper aims to analyze the fast particles scattering in a crystal in this approximation. It was shown that in the Born approximation it is possible to consider easily the contribution of atoms arrangement in matter to the scattering and to examine from a single point of view the fast particles scattering in thin crystals both when particles fall on the crystal along one of its crystallographic axes and planes. The main attention is paid to the analysis of the applicability conditions of the Born approximation in this problem and to the comparative analysis of the scattering characteristics for the different atoms arrangement in the crystal relatively to the incident beam.\par}

\section{DIFFERENTIAL CROSS SECTION OF FAST PARTICLES SCATTERING IN MATTER}

{\fontsize{10pt}{12.0pt}\selectfont Let us consider the fast charged particle scattering at small angles in a thin layer of matter. The potential energy of particle interaction with the atoms of the medium in this case is the sum of the potential energies of its interaction with each atom: \par}

\begin{eqnarray} 
\Centering U \left( \vec{r} \right) = \sum _{n=1}^{N}u \left(\vec{r}-\vec{r}_{n} \right) 
\end{eqnarray} 
 
{\fontsize{10pt}{12.0pt}\selectfont where  \( \vec{r}_{n} \)  is the position of atom in the medium and  \( N \)  is amount of atoms in the medium.\par}

{\fontsize{10pt}{12.0pt}\selectfont The potential energy  \( U \left( \vec{r} \right)  \)  is a complicated coordinate function, depending on the atoms arrangement in the medium, which can be either regular (crystal) or random (amorphous medium). Therefore, to describe the scattering process in such structures, it is important to choose efficient approaches and approximations, which make it possible to carry out calculations in fields of complex configuration. Such methods, particularly, include methods based on the Born and eikonal approximations of the quantum scattering theory, since one does not need to specify the  \( U \left( \vec{r} \right)  \)  function.\par}

{\fontsize{10pt}{12.0pt}\selectfont So, for half-integer spin particles the differential scattering cross section averaged over the polarizations of the initial particles and summed over the polarizations of the final states has the following form in the first Born approximation (in this paper we use the system of units where the light velocity \( c=1 \)) [5]:\par}

\begin{eqnarray} 
\Centering \frac{d \sigma }{do}=\frac{ \varepsilon ^{2}}{4 \pi ^{2}\hslash^{2}} \vert U_{q} \vert ^{2} \left( 1-\frac{\vec{q}^{~2}}{4 \varepsilon ^{2}} \right) 
\end{eqnarray} 
 
{\fontsize{10pt}{12.0pt}\selectfont where  \( do \)  is the solid angle element along the scattering direction,  \(  \varepsilon  \)  is the particle energy,  \( U_{q} \)  is the Fourier component of  \( U \left( \vec{r} \right)  \)  and  \( \vec{q}=\vec{p}-\vec{p}~' \)  is the transmitted momentum to the external field when the particle was scattered. \par}

{\fontsize{10pt}{12.0pt}\selectfont Substituting in (2) the potential energy of the particle interaction with atoms of the substance (1), we obtain the following expression for the scattering cross section:\par}

\begin{eqnarray}
\Centering \frac{d \sigma }{do}= \Bigl|  \sum _{n=1}^{N}e^{i\frac{\vec{q}\vec{r}_{n}}{\hslash}} \Bigr| ^{2}\frac{d \sigma ^{ \left( 1 \right) }}{do}
\end{eqnarray} 
 
{\fontsize{10pt}{12.0pt}\selectfont where  \( \frac{d \sigma ^{ \left( 1 \right) }}{do} \)  is the scattering cross section in the field of a separate atom of the medium \cite{Ahi65}\par}

\begin{eqnarray}
\frac{d \sigma ^{ \left( 1 \right) }}{do}=\frac{ \varepsilon ^{2}}{4 \pi ^{2}\hslash^{2}} \vert u_{q} \vert ^{2} \left( 1-\frac{\vec{q}^{~2}}{4 \varepsilon ^{2}} \right) 
\end{eqnarray} 
 
{\fontsize{10pt}{12.0pt}\selectfont and  \( u_{q} \)  is the Fourier component of the potential energy of the particle interaction with a separate atom of the medium. Thus, in the first Born approximation, the scattering cross section in a substance differs in the diffraction factor \textit{D} from the corresponding cross section of scattering by a single atom of the medium:\par}

\begin{eqnarray} 
\Centering D=  \Bigl|  \sum _{n=1}^{N}e^{i\frac{\vec{q}\vec{r}_{n}}{\hslash}}  \Bigr| ^{2}
\end{eqnarray} 
 
{\fontsize{10pt}{12.0pt}\selectfont If there are no diffraction effects in scattering, then \\
\textit{D = N}, and therefore, \par}

\begin{eqnarray} 
\Centering \frac{d \sigma }{do}=N\frac{d \sigma ^{ \left( 1 \right) }}{do}~.
\end{eqnarray} 
 
{\fontsize{10pt}{12.0pt}\selectfont This situation corresponds to the particles scattering in a rarefied medium, when the atoms are at large distances from each other.\par}

{\fontsize{10pt}{12.0pt}\selectfont At high energies, the particles scattering angles in a substance are typically small compared to unity. Moreover, if scattering occurs in a thin layer of a substance, then under the condition\par}

\begin{eqnarray} 
\Centering q_{l}L \ll 1
\end{eqnarray} 
 
{\fontsize{10pt}{12.0pt}\selectfont where  \( q_{l}=\frac{\vec{q}^{2}}{2p} \)  is the longitudinal component of the transmitted momentum  \( \vec{p} \)  and \textit{L} is the target thickness, in (3) we can neglect the dependence of the scattering cross section on  \( q_{l} \) . In this case, the diffraction effects in scattering are determined only by the distribution of atoms in the target in the plane orthogonal to  \( \vec{p} \) . Moreover, in particular, if all the atoms are located along a line parallel to  \( \vec{p} \) , then, according to (3),\par}

\begin{eqnarray}
\Centering \frac{d \sigma }{do}=N^{2}\frac{d \sigma ^{ \left( 1 \right) }}{do}.
\end{eqnarray} 
 
{\fontsize{10pt}{12.0pt}\selectfont The proportionality of the scattering cross section to the squared number of atoms in the target in this case indicates a coherent scattering effect.\par}

{\fontsize{10pt}{12.0pt}\selectfont The applicability condition for the Born approximation in describing the process of coherent scattering by a string of crystal atoms (8) has the following form \cite{Ahi96}:\par}

\begin{eqnarray} 
\frac{NZe^{2}}{\hslash v} \ll 1
\end{eqnarray} 
 
{\fontsize{10pt}{12.0pt}\selectfont where  \( Z \vert e \vert  \)  is the charge of the nucleus of an atom in the string and  \( v \)  is the particle velocity. This condition is rapidly violated with increasing the number of atoms in the string. In description of the process of particles scattering by a string of atoms, in this case, it is necessary to go beyond the Born approximation. Such description for high energies could be based on the eikonal approximation of the quantum scattering theory. The differential cross section of scattering for unpolarized particles in this approximation has the following form \cite{Ahi96} \par}

\begin{eqnarray} 
\Centering \frac{d \sigma }{do}=\frac{p^{2}}{4 \pi ^{2}}  \Bigl|  \int _{}^{}d^{2} \rho e^{i\frac{\vec{q}_{\bot}\vec{ \rho }}{\hslash}} \left( e^{i\frac{ \chi  \left( \vec{ \rho } \right) }{\hslash}}-1 \right)  \Bigr| ^{2}
\end{eqnarray} 
 
{\fontsize{10pt}{12.0pt}\selectfont where  \( \vec{ \rho }= \left( x,y \right)  \)  are coordinates in orthogonal to  \( \vec{p} \)  plane and  \(  \chi  \left( \vec{ \rho } \right)  \)  is the scattering phase,\par}

\begin{eqnarray} 
\Centering  \chi  \left( \vec{ \rho } \right) =-\frac{1}{v} \int _{-\infty}^{\infty}dzU \left( \vec{ \rho },z \right) 
\end{eqnarray} 
 
{\fontsize{10pt}{12.0pt}\selectfont The formula (10) is valid for fast particles scattering at small angles in a localized field  \( U \left( \vec{r} \right)  \) , provided that the particle motion in this field is close to rectilinear one \cite{Ahi96}. This requires that corrections in the eikonal scattering phase are small. This requirement is satisfied at sufficiently high particle energies, since the noted corrections are proportional to  \( p^{-1} \)  \cite{Ahi96}. As for the magnitude of the scattering phase  \(  \vert  \chi  \left( \vec{ \rho } \right)  \vert  \) , it can be either small or large compared to the Planck constant $\hslash$ .\par}

{\fontsize{10pt}{12.0pt}\selectfont Under the condition:\par}

\begin{eqnarray} 
\Centering  \vert  \chi  \left( \vec{ \rho } \right)  \vert  \ll \hslash
\end{eqnarray} 
 
{\fontsize{10pt}{12.0pt}\selectfont we can expand (10) with the small parameter  \( \frac{ \chi  \left( \vec{ \rho } \right) }{\hslash} \) . In the first non-vanishing approximation of this expansion, formula (10) transforms into the corresponding result of the Born approximation (2). Thus, inequality (12) is an applicability condition of the Born approximation for describing the fast particles scattering in matter.\par}

{\fontsize{10pt}{12.0pt}\selectfont For particles scattering on a string of atoms, condition (12) leads to the inequality (9), which determines the applicability condition of the Born approximation for the problem of fast particles coherent scattering in a thin crystal. When a particle is scattered in an amorphous medium, inequality (12) can be written in the following form:\par}

\begin{eqnarray} 
\Centering \frac{L}{l_{MFP}}\frac{Ze^{2}}{\hslash v} \ll 1
\end{eqnarray} 
 
{\fontsize{10pt}{12.0pt}\selectfont where  \( L \)  is the target thickness and  \( l_{MFP} \)  is the mean free path of a particle in a substance between its successive collisions with atoms. Here,  \( \frac{L}{l_{MFP}} \)  represents the number of collisions of a particle with atoms during the passage of a target of thickness \textit{L}. If condition (13) is violated, it is necessary to consider effects associated with the multiple particle scattering by atoms in an amorphous medium. \par}

\section{BORN APPROXIMATION FOR THE FAST PARTICLES ELASTIC SCATTERING CROSS SECTION IN ORIENTED CRYSTALS}

{\fontsize{10pt}{12.0pt}\selectfont Let us consider the fast charged particles scattering in a thin crystal at small angles as particles fall along one of the crystallographic axes (\textit{z} axis). By the thin crystal we mean a crystal which thickness satisfies condition (7). In this case, the particle scattering cross section in the first Born approximation is determined by formula (3).\par}

{\fontsize{10pt}{12.0pt}\selectfont The positions of atoms in the crystal have a periodic structure with a small positions deviation  \( \vec{u}_{n} \)  of each atom relatively to its equilibrium positions  \( \vec{r}_{n}^{~0}= \left( \vec{ \rho }_{n}^{~0}, z_{n}^{0} \right)  \):\par}

\begin{eqnarray} 
\Centering \vec{r}_{n}=\vec{r}_{n}^{~0}+\vec{u}_{n}.
\end{eqnarray} 
 
{\fontsize{10pt}{12.0pt}\selectfont This spreading of atoms positions is due to atoms thermal vibrations in the lattice and it leads to necessity of averaging the formula for the scattering cross section (3). Let us assume for simplicity hereafter that the  \( \vec{u}_{n} \)  distribution function is Gaussian\par}

\begin{eqnarray} 
\Centering f \left( \vec{u} \right) =\frac{1}{ \left( 2 \pi \overline{u^{2}} \right) ^{\frac{3}{2}}}e^{-\frac{\vec{u}^{2}}{2\overline{u^{2}}}}
\end{eqnarray} 
 
{\fontsize{10pt}{12.0pt}\selectfont with the mean squared amplitude of the thermal vibrations of atoms along each crystallographic axis equal to  \( \overline{u^{2}} \) = \( \overline{u_{x}^{2}}=\overline{u_{y}^{2}}=\overline{u_{z}^{2}} \) . Firstly, let us consider the simplest case of a particle scattering on a string of  \( N_{z} \)  atoms located along the direction of the incident particles momentum, \textit{z}-axis. As a result of averaging over atoms thermal vibrations, in this case we find the following expression for the mean value of the scattering cross section (3) at small angles\par}

\begin{eqnarray} 
 & 
 4 \pi ^{2} \langle \frac{d^{2} \sigma }{dq_{\bot}^{2}} \rangle =N_{z} \left( 1-e^{- \frac{\vec{q}^{2}\overline{u^{2}}}{\hslash^{2}}} \right)  \vert u_{\vec{q}} \vert ^{2} + &  {} \nonumber\\ 
& +N_{z}^{2}e^{- \frac{\vec{q}^{2}\overline{u^{2}}}{\hslash^{2}}} \vert u_{\vec{q}} \vert ^{2}  & \end{eqnarray}

{\fontsize{10pt}{12.0pt}\selectfont The first term in (16) does not depend on the atoms arrangement in strings. This term determines the incoherent effects in scattering. The interference effects in the particle (plane wave) scattering by crystal atoms are determined by the second term in (16).\par}

{\fontsize{10pt}{12.0pt}\selectfont Now we consider the scattering on a set of strings of atoms in the crystal, located periodically in the (\textit{x,y}) plane, orthogonal to the momentum  \( \vec{p} \)  of the incident particles. Later we will consider the simplest version of the atoms distribution in such crystal, corresponding to a crystal with a cubic lattice with distance \textit{a} between atoms along each axis. From the energy and momentum conservation laws it follows that the longitudinal component of the transmitted momentum  \( q_{z} \)  is determined by the relation\par}

\begin{eqnarray} 
\Centering q_{z}=\frac{q_{\bot}^{2}+q_{z}^{2}}{2p}
\end{eqnarray} 
 
{\fontsize{10pt}{12.0pt}\selectfont For a fixed value of the transverse component of the transmitted momentum \( q_{\bot}=\sqrt{q_{x}^{2}+q_{y}^{2}} \) and sufficiently high values of the particle energy, condition (7) is always satisfied. Under this condition we can neglect the dependence of the scattering cross section (16) on  \( q_{z} \) . The diffraction factor in (16) in this case after averaging over atoms thermal vibrations has the following form\par}

\begin{eqnarray} 
\Centering & D=N_{z} \left( 1-e^{-~ \frac{q_{\bot}^{2}\overline{u^{2}}}{\hslash^{2}}} \right) + {} \nonumber\\
\Centering & + N_{z}^{2}  \Bigl|  \sum _{n_{x},n_{y}=1}^{N_{x},N_{y}}e^{i\frac{\vec{q}_{\bot}\vec{ \rho }_{n}^{~0}}{\hslash}}  \Bigr| ^{2}e^{-~ \frac{q_{\bot}^{2}\overline{u^{2}}}{\hslash^{2}}}~. &
\end{eqnarray} 
 
{\fontsize{10pt}{12.0pt}\selectfont where  \( N_{x} \) ,  \( N_{y} \)  and  \( N_{z} \)  are numbers of atoms along \textit{x}, \textit{y} and \textit{z }axes,  \( N=N_{z}N_{x}N_{y} \)  is the total amount of atoms in the crystal. \tabto{6.5in} \par}

{\fontsize{10pt}{12.0pt}\selectfont The quadratic dependence of the diffraction factor on  \( N_{z} \)  leads to a coherent scattering effect in the provided case of the crystal axes orientation relatively to the direction of the incident particles momentum. In particular, for scattering in crystal on its separate string of atoms located strictly on the z axis, the scattering cross section (18) transforms into the corresponding result (16) of the coherent scattering theory for particle scattered by a string of atoms (in this case,  \( N_{x}=N_{y}=1 \)  and  \( \overline{u^{2}} \rightarrow 0 \) ).\par}

{\fontsize{10pt}{12.0pt}\selectfont If the positions of crystal strings axes in the transverse plane form a random structure, then averaging over the positions of the strings of atoms in the transverse plane, we obtain the following expression for the differential scattering cross section in this case\par}

\begin{eqnarray} 
\Centering & \langle \frac{d^{2} \sigma }{dq_{\bot}^{2}} \rangle =\frac{1}{4 \pi ^{2}} \left\{ N \left( 1-e^{- \frac{q_{\bot}^{2}\overline{u^{2}}}{\hslash^{2}}} \right) + \right. &  {} \nonumber\\ 
\Centering & \left. +N_{z}^{2}N_{x}N_{y}e^{- \frac{q_{\bot}^{2}\overline{u^{2}}}{\hslash^{2}}} \right\}  \vert u_{q_{\bot}} \vert ^{2} &
\end{eqnarray} 
 
{\fontsize{10pt}{12.0pt}\selectfont Formula (19) shows that with the random arrangement of crystal strings axes in the \textit{(x, y)} plane, there is no interference effect in scattering on different strings of atoms, while during scattering on each string of atoms the coherent effect is present.\par}

{\fontsize{10pt}{12.0pt}\selectfont When atomic strings are periodically arranged in the transverse plane, summation in the diffraction factor (18) leads to the following result \par}

\begin{eqnarray} 
\Centering D_{coh}=N_{z}^{2}\frac{\sin ^{2} \left( \frac{N_{x}aq_{x}}{2\hslash} \right) }{\sin ^{2} \left( \frac{aq_{x}}{2\hslash} \right) }\frac{\sin ^{2} \left( \frac{N_{y}aq_{y}}{2\hslash} \right) }{\sin ^{2} \left( \frac{aq_{y}}{2\hslash} \right) }
\end{eqnarray} 
 
{\fontsize{10pt}{12.0pt}\selectfont For large values of  \( N_{x} \)  and  \( N_{y} \)  we obtain\  \par}

\begin{eqnarray} 
\Centering D_{coh}=N_{z}^{2}N_{x}N_{y}\frac{ \left( 2 \pi  \right) ^{2}}{a^{2}} \sum_{g}^{} \delta  \left( \frac{\vec{q}_{\bot}-\vec{g}}{\hslash} \right) 
\end{eqnarray} 
 
{\fontsize{10pt}{12.0pt}\selectfont where  \(  \delta  \left( \vec{q}_{\bot}-\vec{g} \right)  \)  two-dimensional Dirac delta function and  \( \vec{g}= \left( g_{x},g_{y} \right) =\frac{2 \pi }{a} \left( n_{x},n_{y} \right)  \)  is reciprocal lattice vector. The differential scattering cross section in this case has the following form\par}

\begin{eqnarray} 
\Centering &  \langle \frac{d^2 \sigma }{dq_{\bot}^{2}} \rangle =\frac{N}{4 \pi ^{2}} \left\{  \left( 1-e^{- \frac{q_{\bot}^{2}\overline{u^{2}}}{\hslash^{2}}} \right) + \right.  &  {} \nonumber\\ 
\Centering & \left.   +N_{z}\frac{ \left( 2 \pi  \right) ^{2}}{a^{2}} \sum_{g}^{} \delta  \left( \frac{\vec{q}_{\bot}-\vec{g}}{\hslash} \right) e^{- \frac{q_{\bot}^{2}\overline{u^{2}}}{\hslash^{2}}} \right\}  \vert u_{q_{\bot}} \vert ^{2} &
\end{eqnarray} 
 
{\fontsize{10pt}{12.0pt}\selectfont It follows from formula (22) that in addition to the coherent effect of scattering on each string of atoms in this case, there is also an interference effect of particle (plane wave) scattering on different strings of atoms.Due to the interference effect, the transmitted pulse transverse component has discrete values \( \vec{q}_{\bot}=\vec{g} \) . In the term in (22), which determines incoherent effects in scattering, there is no such interference effect.\par}

{\fontsize{10pt}{12.0pt}\selectfont Thus, the differential cross section for fast charged particles scattering in a thin crystal substantially depends on the arrangement of atoms and groups of atoms in the target. In this case, when particles move along one of the crystallographic axes, both coherent effect in scattering by crystal strings of atoms and interference effect associated with scattering by different strings of atoms are possible. Considering the thermal spread of atoms positions in the lattice leads to a splitting of the cross section into the sum of coherent and incoherent cross sections. The incoherent scattering cross section does not depend on the location of atoms in the target and slightly differs from the corresponding scattering cross section in an amorphous medium.\par}

\section{SCATTERING BY CRYSTALLINE PLANES OF ATOMS}

\vspace{\baselineskip}
 
{\fontsize{10pt}{12.0pt}\selectfont Of particular interest is the case of the fast charged particles incidence on a crystal along one of the crystallographic planes, since the scattering of particles in different directions are in this case of different types. In this connection, we consider scattering by a system of crystalline planes periodically arranged along the \textit{x}-axis (see Fig. 1), assuming for simplicity that atoms positions in each of planes are equally probable (the \textit{x}-axis is perpendicular to the crystalline planes of atoms \textit{(y,z)}).\par}

{\fontsize{10pt}{12.0pt}\selectfont The averaging procedure of the scattering cross section (3) in this case is connected to the diffraction factor (5), in which the position components  \( \vec{r}_{n}= \left( x_{n},y_{n},z_{n} \right)  \) \  of each atom have the following form\par}

\begin{eqnarray} 
x_{n}=an_{x}+u_{n_{x},n_{p}}, y_{n}=y_{n_{x},n_{p}}, z_{n}=z_{n_{x},n_{p}},
\end{eqnarray}

{\fontsize{10pt}{12.0pt}\selectfont where index  \( n_{x} \)  is a plane of atoms index number,  \( n_{p} \)  is atom index number in the plane and   \( u_{n_{x},n_{p}} \)  is thermal deviation of  \( n_{p} \) -th atom in  \( n_{x} \) -th plane from the \textit{x}-axis.\par}

\vspace{\baselineskip}


\begin{table}[H]
 			\centering
\begin{tabular}{p{2.94in}}
\multicolumn{1}{p{2.94in}}{
	\begin{Center}
		\includegraphics[width=2.94in,height=2.36in]{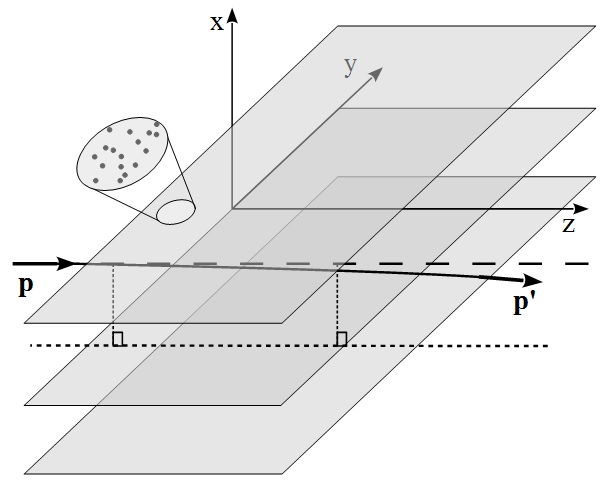}
	\end{Center}
} \\
\hhline{~}
\multicolumn{1}{p{2.94in}}{{\fontsize{10pt}{12.0pt}\selectfont \textbf{\textit{Fig. 1.}}\textit{ Scattering by periodic planes of atoms with homogeneous distribution of atoms}}} \\
\hhline{~}

\end{tabular}
 \end{table}


\vspace{\baselineskip}
 
{\fontsize{10pt}{12.0pt}\selectfont Using (23) we obtain the following expression for diffraction factor D:\par}

\begin{eqnarray} 
\Centering & D= \Bigl|  \sum_{n_{x}=1}^{N_{x}} e^{i\frac{q_{x} an_{x}}{\hslash}}   \times  &  {} \nonumber\\ 
&  \times  \sum_{n_{p}=1}^{N_{p}} \exp  \left[ i\frac{q_{x}u_{n_{x},n_{p}}+q_{y}y_{n_{x},n_{p}}+q_{z}z_{n_{x},n_{p}}}{\hslash} \right]  \Bigr| ^{2}, &
\end{eqnarray} 
 
{\fontsize{10pt}{12.0pt}\selectfont where  \( N_{x} \)  is the number of crystalline planes arranged along \textit{x}-axis and  \( N_{p} \)  is the number of atoms in each plane (it is assumed that the numbers of atoms in each plane are the same).\par}

{\fontsize{10pt}{12.0pt}\selectfont Due the equiprobability of the positions of atoms in each plane and assuming that the law of thermal displacement of each atom along the x axis has a Gaussian form with the same squared displacement along this axis equal to  \( \overline{u^{2}} \) , we arrive at the following expression for the average value of the diffraction factor \textit{D}: \par}

\begin{eqnarray} 
\Centering & \langle D \rangle = \prod_{k_{x}=1}^{N_{x}} \prod_{k_{p}=1}^{N_{p}} \int _{-\infty}^{\infty}du_{x,k_{x},k_{p}}f \left( u_{x,k_{x},k_{p}} \right)  \times  &  {} \nonumber\\ 
\Centering &   \times  \int _{-\frac{L_{y}}{2}}^{\frac{L_{y}}{2}}\frac{dy_{k_{x},k_{p}}}{L_{y}} \int _{-\frac{L_{z}}{2}}^{\frac{L_{z}}{2}}\frac{dz_{k_{x},k_{p}}}{L_{z}}D, &
\end{eqnarray} 
 
{\fontsize{10pt}{12.0pt}\selectfont where  \( L_{y} \)  and  \( L_{z} \)  are crystalline plane sizes along  \( y \) - and  \( z \) -axis and\par}

\begin{eqnarray} 
\Centering  f \left( u_{x,k_{x},k_{p}} \right) = \frac{1}{\sqrt{2 \pi \overline{u^{2}}}}\exp  \left( -\frac{u_{x,k_{x},k_{p}}^{2}}{2\overline{u^{2}}} \right) . 
\end{eqnarray} 
 
{\fontsize{10pt}{12.0pt}\selectfont The integrals in (25) in each term have the following structure if  \( n_{x} \neq n_{x}^{'} \)  and if  \( n_{x}=n_{x}^{'} \) ,  \( n_{p} \neq n_{p}^{'} \) \par}

\begin{eqnarray} 
\Centering & \int _{-\frac{L_{y}}{2}}^{\frac{L_{y}}{2}}\frac{dy_{n_{x},n_{p}}}{L_{y}} \int _{-\frac{L_{y}}{2}}^{\frac{L_{y}}{2}}\frac{dy_{n_{x}^{'},n_{p}^{'}}}{L_{y}}e^{i\frac{q_{y} \left( y_{n_{x},n_{p}}-y_{n_{x}^{'},n_{p}^{'}} \right) }{\hslash}}=  &  {} \nonumber\\ 
& =\left( \frac{\sin  \left( \frac{q_{y}L_{y}}{2\hslash} \right) }{\frac{q_{y}L_{y}}{2\hslash}} \right) ^{2} &
\end{eqnarray} 
 
{\fontsize{10pt}{12.0pt}\selectfont If  \( n_{x}=n_{x}^{'} \)  and  \( n_{p}=n_{p}^{'} \) , then this integrals are equal to unity. \par}

{\fontsize{10pt}{12.0pt}\selectfont The integrals related to averaging over  \( z_{n_{x},n_{p}} \)  and  \( z_{n_{x}^{'},n_{p}^{'}} \)  have a similar structure. However, in this case under the condition  \( \frac{q_{z}L_{z}}{2} \ll \hslash \)   all terms of averaging over these variables are equal to unity.\par}

{\fontsize{10pt}{12.0pt}\selectfont For  \( n_{x} \neq n_{x}^{'} \)  and for  \( n_{x}=n_{x}^{'} \) ,  \( n_{p} \neq n_{p}^{'} \)  integrals of averaging over the thermal vibrations of  \( u_{x} \)  atoms have the following form\par}

\begin{eqnarray} 
\Centering &  \int _{-\frac{L_{y}}{2}}^{\frac{L_{y}}{2}}du_{x,n_{x},n_{p}}f \left( u_{x,n_{x},n_{p}} \right)  \times  &  {} \nonumber\\ 
\Centering  &  \int _{-\frac{L_{y}}{2}}^{\frac{L_{y}}{2}}du_{x,n_{x}^{'},n_{p}^{'}}f \left( u_{x,n_{x}^{'},n_{p}^{'}} \right) e^{i\frac{q_{x} \left( u_{x,n_{x},n_{p}}-u_{x,n_{x}^{'},n_{p}^{'}} \right) }{\hslash}} &  {} \nonumber\\ 
& = e^{- \frac{q_{\bot}^{2}\overline{u^{2}}}{\hslash^{2}}}  &
\end{eqnarray} 
 
{\fontsize{10pt}{12.0pt}\selectfont If  \( n_{x}=n_{x}^{'} \)  and  \( n_{p}=n_{p}^{'} \) , then these integrals are equal to unity.\par}

{\fontsize{10pt}{12.0pt}\selectfont Substituting obtained averaging results in (25), we find that\par}

\begin{eqnarray} 
  \langle D \rangle = \sum _{n_{x}=n_{x}^{'}}^{N_{x}} \sum _{n_{p}=n_{p}^{'}}^{N_{p}}1+A \sum _{n_{x}=n_{x}^{'}}^{N_{x}} \sum _{n_{p} \neq n_{p}^{'}}^{N_{p}}1+ &  {} \nonumber\\ 
+A \sum _{n_{x} \neq n_{x}^{'}}^{N_{x}}e^{i\frac{q_{x}a \left( n_{x}-n_{x}^{'} \right) }{\hslash}} \sum _{n_{p},n_{p}^{'}=1}^{N_{p}}1, &
\end{eqnarray} 
 
{\fontsize{10pt}{12.0pt}\selectfont where\par}

\begin{eqnarray} 
\Centering A=e^{- \frac{q_{\bot}^{2}\overline{u^{2}}}{\hslash^{2}}} \left( \frac{\sin  \left( \frac{q_{y}L_{y}}{2\hslash} \right) }{\frac{q_{y}L_{y}}{2\hslash}} \right) ^{2}.
\end{eqnarray} 
 
{\fontsize{10pt}{12.0pt}\selectfont Adding in (29) to terms containing  \( n_{p} \neq n_{p}^{'} \)  and  \( n_{x} \neq n_{x}^{'} \)  summands, terms with  \( n_{p}=n_{p}^{'} \)  and, respectively,  \( n_{x}=n_{x}^{'} \)  and subtracting similar terms, we obtain (29) in the form\par}

\begin{eqnarray} 
\Centering &  \langle D \rangle =N_{x}N_{p}+AN_{x} \left(  \sum _{n_{p},n_{p}^{'}}^{N_{p}}1-N_{p} \right) +  &  {} \nonumber\\ 
&+AN_{p}^{2} \left(  \sum _{n_{x},n_{x}^{'}=1}^{N_{x}}e^{i\frac{q_{x}a \left( n_{x}-n_{x}^{'} \right) }{\hslash}}-N_{x} \right)  &
\end{eqnarray} 
 
{\fontsize{10pt}{12.0pt}\selectfont As a result, we find that\par}

\begin{eqnarray} 
\Centering  \langle D \rangle =N_{x}N_{p} \left( 1-A \right) +N_{p}^{2}A\frac{\sin ^{2} \left( \frac{N_{x}q_{x}a}{2\hslash} \right) }{\sin ^{2} \left( \frac{q_{x}a}{2\hslash} \right) }
\end{eqnarray} 
 
{\fontsize{10pt}{12.0pt}\selectfont For large  \( L_{y} \)  and  \( N_{x} \)  values we have\par}

\begin{eqnarray} 
\Centering & A=e^{- \frac{q_{\bot}^{2}\overline{u^{2}}}{\hslash^{2}}}\frac{2 \pi }{L_{y}} \delta  \left( \frac{q_{y}}{\hslash} \right) , &  {} \nonumber\\ 
\Centering  & \frac{\sin ^{2} \left( \frac{N_{x}q_{x}a}{2\hslash} \right) }{\sin ^{2} \left( \frac{q_{x}a}{2\hslash} \right) }=N_{x}\frac{2 \pi }{a} \sum _{g}^{} \delta  \left( \frac{q_{x}-g}{\hslash} \right) , &
\end{eqnarray} 
 
{\fontsize{10pt}{12.0pt}\selectfont where  \( g=\frac{2 \pi n}{a} \)  and  \( n=0, \pm 1,  \ldots  \)  . Substituting (33) in (32) results in\par}

\begin{eqnarray} 
\Centering &  \langle D \rangle =N_{x}N_{p} \left( 1-\frac{2 \pi }{L_{y}} \delta  \left( \frac{q_{y}}{\hslash} \right) e^{- \frac{q_{\bot}^{2}\overline{u^{2}}}{\hslash^{2}}} \right) + &  {} \nonumber\\ 
\Centering & +N_{x}N_{p}^{2}\frac{2 \pi }{L_{y}} \delta  \left( \frac{q_{y}}{\hslash} \right) \frac{2 \pi }{a} \sum _{g}^{} \delta  \left( \frac{q_{x}-g}{\hslash} \right) e^{- \frac{q_{x}^{2}\overline{u^{2}}}{\hslash^{2}}}. &
\end{eqnarray} 
 
{\fontsize{10pt}{12.0pt}\selectfont Noting that the total number of atoms in the crystal is  \( N=N_{x}N_{p} \)  and that the number of atoms in each plane is  \( N_{p}=n_{\bot}L_{y}L_{z} \) , where  \( n_{\bot} \)  is the density of atoms in a separate plane, we find that if  \( L_{y} \rightarrow \infty \) \  the scattering cross section (3) has the following form\par}

\begin{eqnarray} 
\Centering & \langle \frac{d^{2} \sigma }{dq_{\bot}^{2}} \rangle =N \left\{ 1+n_{\bot} L_{z}\frac{ \left( 2 \pi  \right) ^{2}}{a} \delta  \left( \frac{q_{y}}{\hslash} \right)   \times   \right.  &  {} \nonumber\\ 
\Centering & \left. \times  \sum _{g}^{} \delta  \left( \frac{q_{x}-g}{\hslash} \right) e^{- \frac{q_{x}^{2}\overline{u^{2}}}{\hslash^{2}}} \right\}  \langle \frac{d^{2} \sigma ^{ \left( 1 \right) }}{dq_{\bot}^{2}} \rangle &
\end{eqnarray} 
 
{\fontsize{10pt}{12.0pt}\selectfont The first term in braces determines the incoherent effects in scattering. This term is proportional to the number of atoms in the crystal and completely coincides with the corresponding result of the scattering theory for amorphous medium. The second term determines the coherent and interference effect in scattering. This term has an additional factor proportional to the thickness of the crystal, due to which the scattering of particles by the crystal is intensified compared to scattering in amorphous medium. This intensification of scattering is connected to correlations in particle collisions with atoms of separate crystalline planes of atoms. The number of such collisions and the corresponding intensification in the case is in the order of  \( K \sim L_{z}R/a^{2} \) .\par}

{\fontsize{10pt}{12.0pt}\selectfont Delta function  \(  \delta  \left( q_{x}-g \right)  \)  in the second term in (35) is due to the interference effect in scattering by different crystal planes arranged periodically along the \textit{x}-axis.\par}

{\fontsize{10pt}{12.0pt}\selectfont In the case of particle scattering on the crystal planes of atoms, there is no Debye-Waller factor in the term that determines incoherent effects in scattering.\par}

{\fontsize{10pt}{12.0pt}\selectfont We note that formula (35) can also be obtained from formula (22) if, deriving the latter, we consider different deviations of the atoms positions in the \textit{x} and \textit{y} directions and formally set the value of this deviation along the y-axis as  \( \sqrt{\overline{u_{y}^{2}}} \) , which tends to infinity. Moreover, in (22), we should replace the Debye-Waller factor  \( e^{- \frac{q_{\bot}^{2}\overline{u^{2}}}{\hslash^{2}}} \)  with the  \( e^{- \frac{q_{x}^{2}\overline{u_{x}^{2}}}{\hslash^{2}}}e^{- \frac{q_{y}^{2}\overline{u_{y}^{2}}}{\hslash^{2}}}~ \) and use the relation \cite{Gel64} \par}

\begin{eqnarray} 
\Centering e^{- \frac{q_{y}^{2}\overline{u_{y}^{2}}}{\hslash^{2}}} \approx \frac{\hslash\sqrt{ \pi }}{\sqrt{\overline{u_{y}^{2}}}} \delta  \left( q_{y} \right) 
\end{eqnarray} 
 
{\fontsize{10pt}{12.0pt}\selectfont which can be used if  \( \sqrt{\overline{u_{y}^{2}}} \rightarrow \infty \) .\par}

{\fontsize{10pt}{12.0pt}\selectfont Then the last term in (22) is proportional to the product of the delta-functions:\par}

\begin{eqnarray} 
\Centering \frac{1}{L_{y}} \sum _{q_{y}}^{} \delta  \left( q_{y}-g_{y} \right)  \delta  \left( q_{y} \right) =\frac{ \delta  \left( q_{y} \right) }{2 \pi }
\end{eqnarray} 

\vspace{\baselineskip}

\section{DISCUSSION}

\vspace{\baselineskip}
 
{\fontsize{10pt}{12.0pt}\selectfont The obtained results indicate that, in the Born approximation of quantum theory, the cross section of elastic scattering of fast charged particles in a thin crystal splits into cross sections of coherent and incoherent scattering. The coherent scattering cross section determines interference effects in the scattering of a particle by numerous atoms of a crystal. This cross section significantly depends on the orientation of the crystallographic axes and planes with respect to the motion direction of the particles incident on the crystal. Moreover, if the particle passes the crystal along one of the crystalline axes, then on the condition\par}

\begin{eqnarray} 
L \ll \hslash q_{z}^{-1}=\frac{2\hslash p}{q_{\bot}^{2}}
\end{eqnarray} 
 
{\fontsize{10pt}{12.0pt}\selectfont there is a coherent effect in scattering by strings of crystal atoms located along this axis. This effect is manifested in the quadratic dependence of the scattering cross section on the number of atoms in a string. The scattering cross section in this case is in fact determined by the continuous potential of the atomic strings of the crystal, i.e. by the lattice potential averaged over z-axis, which is widely used in the theory of the axial channeling phenomenon in a crystal \cite{Lin}. Thus, the concept of the continuous potential of atomic strings of a crystal naturally appears in the Born theory of particles scattering in thin crystals, that is, under conditions when the channeling phenomenon is absent.\par}

{\fontsize{10pt}{12.0pt}\selectfont Accounting the periodicity of atoms strings axes in a crystal in the transverse plane leads to an interference effect in the scattering of a particle (plane wave) by different strings of atoms, this effect consists in that the transverse components of the transmitted momentum are equal to corresponding components of reciprocal lattice vector, multiplied by integer values. However, if positions of the strings axes in the transverse plane can be formally considered random (this situation corresponds to the conditions for the dynamic chaos occurrence during particle motion in the crystal \cite{Akh91}), there is no interference effect in scattering and the values of the components of the transmitted momentum \( \vec{q}_{\bot} \)  can be arbitrary.\par}

{\fontsize{10pt}{12.0pt}\selectfont The incoherent scattering cross section does not depend on the orientation of the crystal axes relatively to the incident beam. This cross section, however, differs a bit from the corresponding cross section for particle scattering in an amorphous medium. The difference is due to the presence in the cross section of an additional term containing the Debye-Waller factor  \( \exp  \left( -\vec{q}_{\bot}^{2}\overline{u^{2}} \right)  \) . With this term, the incoherent scattering cross section is about 10$\%$  smaller than the corresponding cross section in an amorphous medium. For  \( \overline{u^{2}} \rightarrow \infty \) , this addend to the scattering cross section, as well as the coherent scattering cross section disappear, and the scattering cross sections of particles in an amorphous medium and in a crystal coincide.\par}

{\fontsize{10pt}{12.0pt}\selectfont A similar situation with the splitting of the scattering cross section of fast charged particles in a crystal into coherent and incoherent components is also possible with the passage of particles along thin (longitudinally) crystalline planes of atoms. In this case, however, new variations of scattering appear due to the different origin of the atoms distribution in the crystalline planes of atoms (regular and random) and the presence of periodically arranged atomic planes. Moreover, as shown in the work, for a random arrangement of atoms in each plane in the incoherent scattering cross section, there is no term containing the Debye – Waller factor and this cross section coincides with the corresponding cross section for the amorphous medium. The coherent scattering cross section corresponds to the scattering cross section in the field of the continuous potential of the crystal planes of atoms (since the momentum transferred with  \( q_{y}=0 \)  contributes to this cross section). We note that the crystalline planes of atoms consist, generally speaking, of crystal strings of atoms, located in these planes parallel to each other. In this case, however, if the particles incidents on the crystal along the crystal strings of atoms at large angles to these strings (angles about\\
 \( 1> \psi \gtrsim\frac{R}{a} \) ), then the correlations between successive collisions of the particles with the atoms of the strings are destroyed. Collisions of a particle with different atoms of the plane in this case can be considered as random. This model of the particle interaction with atoms of the plane significantly simplifies calculations.\par}

\section*{Acknowledgements}

{The work was partially supported by the projects L10/56-2019 and C-2/50-2018 of the National Academy of Science of Ukraine. \par}
\vspace{\baselineskip}

\end{multicols}

\end{document}